\documentclass[aps,pra,showpacs,twocolumn]{revtex4}
\begin{document}

\title
{Minimum-error discrimination between
 subsets of linearly dependent quantum states}

\author{Ulrike Herzog$^1$}
\author{J\'{a}nos A. Bergou$^{2,3}$}
\affiliation{$^1$Institut f\"ur Physik,  Humboldt-Universit\"at zu
    Berlin, Invalidenstrasse 110, D-10115 Berlin, Germany}
\affiliation{$^2$Department of Physics, Hunter College, City
    University of New York, 695 Park Avenue, New York, NY 10021,
    USA}
\affiliation{$^3$Institute of Physics, Janus Pannonius University,
    H-7624 P\'{e}cs, Ifj\'{u}s\'{a}g \'{u}tja 6, Hungary}

\date{\today}
\begin{abstract}
A measurement strategy is developed for a new kind of hypothesis
testing. It assigns, with minimum probability of error, the state
of a quantum system to one or the other of two complementary
subsets of a set of $N$ given non-orthogonal quantum states
occurring with given a priori probabilities. A general analytical
solution is obtained for $N$ states that are restricted to a
two-dimensional subspace of the Hilbert space of the system.
The result for the special case of three arbitrary but linearly
dependent states is applied to a variety of sets of three
states that are symmetric and equally probable. It is found that,
in this case, the minimum error probability for distinguishing one
of the states from the other two is only about half as large as
the minimum error  probability for distinguishing all three states
individually.
\end{abstract}


\maketitle

\section{Introduction and basic equations}
Due to their nonvanishing mutual overlaps, non-orthogonal quantum
states cannot be perfectly distinguished.  However, stimulated by
the rapid developments in quantum information theory
\cite{nielsen}, the question as to how to discriminate between
non-othogonal states in an optimum way has gained renewed interest
\cite{chefrev}. In particular, in quantum communication protocols
several secure schemes have been suggested based on communicating
via non-orthogonal quantum states. As a result, optimum
discrimination between them became an inherent part of these
schemes. For studying state discrimination, it is assumed that a
quantum system is prepared in one of the $N$ pure states,
$|\psi_{k}\rangle$, that belongs to a given set of non-orthogonal
states, $\{|\psi_1\rangle, |\psi_2\rangle, \ldots
|\psi_N\rangle\}$, and that the {\it a priori} probabilities $\eta_k$
for the preparation of either one of the states $|\psi_k\rangle$
are also known. In order to devise an optimum state-discriminating
measurement, strategies have been developed with respect to
various criteria \cite{chefrev,barn}. The earliest and simplest of
these criteria is the requirement that the probability of getting
a wrong result be as small as possible, with inconclusive results
being forbidden and all states being individually distinguished. A
minimum-error strategy of this kind has been developed for the
case when only two states are given \cite{hel} and for {\it specific}
$N$ state problems \cite{yuen,ban,sasaki1,barnett}, including $N$
symmetric \cite{ban} and multiply symmetric \cite{barnett} states.
Recently the optimum strategy has also been found for three states
exhibiting a mirror-symmetry \cite{andersson} but still
no exact solution has been known for $N > 2$ {\it arbitrary} states.
Using the polarization states
of a single photon, minimum-error discrimination has been
experimentally realized for up to four symmetric non-orthogonal
states \cite{clarke2}.

In this paper we are concerned with a minimum-error strategy that
involves $N> 2$ {\it arbitrary} linearly dependent quantum states,
by considering the following problem: We want to devise a
measurement that allows us to decide, with the smallest possible
error and without inconclusive answers, whether the actual state
of the system belongs to the subset of states $\{|\psi_1\rangle,
\ldots |\psi_M\rangle\}$, or to the complementary subset of the
remaining states $\{|\psi_{M+1}\rangle, \ldots |\psi_N\rangle\}$
with $M<N$.
For three given states the task reduces to distinguishing
the state $|\psi_1\rangle$ from the set of states
$\{|\psi_2\rangle, |\psi_3\rangle\}$ and can be referred to as
quantum state filtering with respect to the state
$|\psi_1\rangle$. This task has recently been investigated for the
particular optimization strategy  that yields unambiguous
discrimination at the expense of allowing inconclusive results to
occur, the probability of which is minimized \cite{sun}.

To treat our general minimum-error problem, we follow the standard
lines and introduce two positive Hermitian quantum detection
operators, $\Pi_0$  and $\Pi_1$ \cite{chefrev,hel}.
We define the operator $\Pi_1$ by the
property that $\langle\psi_k|\Pi_1|\psi_k\rangle$
accounts for the probability to infer, from performing
the measurement, the system to be in one of the states
$\{|\psi_1\rangle, \ldots |\psi_M\rangle\}$,
if it has been prepared in the state
$|\psi_k\rangle$. Obviously, this inference is incorrect if $k>M$.
Similarly, given again the preparation of the state
$|\psi_k\rangle$, the quantity
$\langle\psi_k|\Pi_0|\psi_k\rangle$
denotes the probability for inferring the state of the
system to belong to the subset of states
$\{|\psi_{M+1}\rangle, \ldots |\psi_N\rangle\}$,
which is an erroneous result if $k\leq M$.
Clearly, the relation
\begin{equation}
\Pi_0  + \Pi_1 = \hat{1}
\label{1}
\end{equation}
has to be obeyed, where  $\hat{1}$ is the unit operator.
From the definition of the detection operators it
follows that the probability to get a correct result
reads
\begin{equation}
P^{M(N)}  =   \sum_{k=1}^M \eta_k \langle\psi_k|\Pi_1|\psi_k\rangle
    + \sum_{k=M+1}^N \eta_k \langle\psi_k|\Pi_0|\psi_k\rangle.
\label{2}
\end{equation}
In order to devise the desired minimum-error
measurement scheme, we have to determine
the particular detection operators $\Pi_0$ and $\Pi_1$
that maximize the right-hand side of Eq. (\ref{2})
under the constraint (\ref{1}).
In general, the error-minimizing optimization
problem is a highly nontrivial task.

\section{Solution in two dimensions}

To enable simple analytical solutions, we restrict ourselves
to the case when the $N$ linearly dependent states
span only a two-dimensional Hilbert space.
We note that for three linearly dependent states this is always the
case. First, we show that in the two-dimensional case
it is possible to represent the two detection operators $\Pi_1$ and
$\Pi_0$ by two projection operators onto orthonormal
states $|\mu\rangle$
and $|\nu\rangle$, respectively. To see this,
we start from the expression
$\Pi_1= \lambda_1 |v_1\rangle \langle v_1|+
\lambda_2 |v_2\rangle \langle v_2|$,
with $|v_1\rangle$ and $|v_2\rangle$
being the orthonormal eigenstates
that belong to some non-negative eigenvalues
$\lambda_1$ and $\lambda_2$.
Expanding a particular state $|\psi\rangle$ as
$|\psi\rangle=
\cos \beta|v_1\rangle + \sin \beta|v_2\rangle$,
where a possible relative phase factor has been
included into the definition of $|v_2\rangle$,
we arrive at
\begin{equation}
\langle\psi|\Pi_1|\psi\rangle = |\langle\mu|\psi\rangle|^2,
\label{3}
\end{equation}
provided that we define
$|\mu\rangle = \sqrt{\lambda_1}|v_1\rangle \pm
i \sqrt{\lambda_2}|v_2\rangle$.
With the help of the relation
$\hat{1} = |v_1\rangle \langle v_1|+|v_2\rangle \langle v_2|$,
we obtain in the same way the representation
\begin{equation}
\langle\psi|\Pi_0|\psi\rangle
 = \langle\psi|\hat{1} - \Pi_1|\psi\rangle =
|\langle\nu|\psi\rangle|^2,
\label{4}
\end{equation}
provided that
$|\nu\rangle = \sqrt{1-\lambda_1}|v_1\rangle \pm
i \sqrt{1-\lambda_2}|v_2\rangle$.
Now we require that
$|\langle\mu|\psi_k\rangle|^2 +  |\langle\nu|\psi_k\rangle|^2 = 1$
for an arbitrary state, $|\psi_k\rangle =  \cos \beta_k|v_1\rangle +
{\rm e}^{i\gamma_k}\sin \beta_k|v_2\rangle$,
which implies that $|\mu\rangle\langle\mu| +
|\nu\rangle\langle\nu| = \hat{1}$ has to be fulfilled.
This only holds true when in the representations of
$|\mu\rangle$ and $|\nu\rangle$  opposite signs
are chosen and when  in addition $\lambda_2 = 1 - \lambda_1$,
leading to the orthonormality conditions
$\langle\mu|\mu\rangle=\langle\nu|\nu\rangle=1$ and
$\langle\mu|\nu\rangle=0$. Therefore,
in a two-dimensional Hilbert space the optimization
problem posed by Eqs. (\ref{1}) and (\ref{2}) can
be reduced to the problem of finding the specific
normalized state $|\mu\rangle$ that maximizes the expression
\begin{equation}
P^{M(N)}  =   \sum_{k=1}^M \eta_k |\langle\mu|\psi_k\rangle|^2
    + \sum_{k=M+1}^N \eta_k\,(1-|\langle\mu|\psi_k\rangle|^2) ,
\label{5}
\end{equation}
which follows when $\Pi_1 = |\mu\rangle\langle\mu|$ and $\Pi_0 =
\hat{1} - |\mu\rangle\langle\mu|$ are substituted into Eq.
(\ref{2}). Comparing this to the spectral representation of the
detection operators, introduced before Eq. (\ref{3}), we are led
to identify $|\mu\rangle$ with $|v_{1}\rangle$ and $|\nu\rangle$
with $|v_{2}\rangle$ since the representation is unique. Then
$\lambda_{1}=1$ and $\lambda_{2}=0$ follows. Once the optimum
detection state is known, the maximum achievable probability of
correctly assigning a quantum state to one of the two subsets, as
well as the two detection operators necessary to perform the
optimized measurement, are uniquely determined.

To solve the optimization problem, it is convenient to
write the overlaps between the given states as
\begin{equation}
 \langle\psi_k|\psi_l\rangle \equiv A_{kl}
 = |A_{kl}| {\rm e}^{i\alpha_{kl}} ,
\label{6}
\end{equation}
 and to introduce the auxiliary state vector
\begin{equation}
 |v\rangle =
\frac{1}{\sqrt{1-|A_{12}|^2}}
(\,|\psi_2\rangle -A_{12}|\psi_1\rangle\,).
\label{7}
\end{equation}
For Eq. (\ref{5}) to be valid, we have to assume
that all $N$ given states lie in a two-dimensional
subspace, spanned by the
states $|\psi_1\rangle$ and $|\psi_2\rangle$,
or $|\psi_1\rangle$ and $|v\rangle$,
respectively. Since $\langle v|v\rangle = 1$ and
$\langle \psi_1|v\rangle = 0$,
the states $|\psi_1\rangle$ and $|v\rangle$
provide a suitable orthonormal basis for representing any state,
$|\psi_k\rangle$, as
\begin{equation}
|\psi_k\rangle = A_{1k}|\psi_1\rangle
      + {\rm e}^{i\gamma_k}
      \sqrt{1-|A_{1k}|^2}\,|v\rangle ,
\label{8}
\end{equation}
with
\begin{equation}
{\rm e}^{i\gamma_k} = \frac{A_{2k} - A_{21}A_{1k}}
{\sqrt{1-|A_{12}|^2} \sqrt{1-|A_{1k}|^2}}.
\label{9}
\end{equation}
The last equation can be verified by calculating the overlap
$\langle\psi_2|\psi_k\rangle$, taking into account that
$\gamma_2= 0$ because of the specific definition of
the state $|v\rangle$.
Similarly, we represent the detection state, $|\mu\rangle$, as
\begin{equation}
|\mu\rangle = \cos\varphi \,|\psi_1\rangle
      + {\rm e}^{i\chi} \sin \varphi \, |v\rangle ,
\label{10}
\end{equation}
and obtain
\begin{equation}
\langle\mu|\psi_k\rangle=
 A_{1k}\cos\varphi
      + {\rm e}^{i(\gamma_k-\chi)} \sqrt{1-|A_{1k}|^2}
      \sin \varphi.
\label{11}
\end{equation}
Eq. (\ref{10}) accounts for all possible states in the
two-dimensional Hilbert space of interest
provided that both $\varphi$ and  $\chi$ are
variables in the interval $[0,\pi)$.
The error-minimization problem is then reduced to finding those
values of $\varphi$ and $\chi$ in Eq. (\ref{11}) that maximize the
probability $P^{M(N)}(\varphi,\chi)$ in Eq. (\ref{5}).

The solution to this optimization problem is straightforward. We
begin by inserting Eq. (\ref{11}) into Eq. (\ref{5}) and, by
making use of the fact that the {\it a priori} probabilities of
the states fulfill the relation $\sum_{k=1}^N \eta_k =1$, we
readily arrive at
\begin{equation}
P^{M(N)} =
   \frac{1}{2}  + R \cos (2\varphi) + |Q| \sin (2\varphi)
   \cos(\chi - \chi_{\rm Q}) ,
\label{12}
\end{equation}
where $R$ and $Q$ are defined as
\begin{equation}
R = \sum_{k=1}^M \eta_k \left(|A_{1k}|^2-\frac{1}{2}\right)
         -\sum_{k=M+1}^N \eta_k \left(|A_{1k}|^2-\frac{1}{2}\right) ,
\label{15}
\end{equation}
and
\begin{eqnarray}
Q &\equiv& |Q|  {\rm e}^{i\chi_{\rm Q}} = \sum_{k=1}^M \eta_k
\frac{A_{2k}
A_{k1}- A_{21} |A_{1k}|^{2}} {\sqrt{1-|A_{12}|^2}} \nonumber \\
       {} &{}& - \sum_{k=M+1}^N \eta_k \frac{A_{2k} A_{k1} -
A_{21} |A_{1k}|^{2}} {\sqrt{1-|A_{12}|^2}}. \label{16}
\end{eqnarray}
The conditions for an extremum, $\partial P^{M(N)}/\partial
\varphi = 0$ and $\partial P^{M(N)}/\partial \chi = 0$, hold for
$\varphi=\varphi_{\rm e}$ and $\chi = \chi_{\rm e}$, with
$\sin(2\varphi_{\rm e}) = |Q|/\sqrt{R^{2}+|Q|^{2}}$,
$\cos(2\varphi_{\rm e}) = R/\sqrt{R^{2}+|Q|^{2}}$, and $\chi_{\rm
e} = \chi_{\rm Q}$, respectively. Note that $\cos(2\varphi_{\rm
e})$ and $R$ have the same sign while $\sin(2\varphi_{\rm e})$ is
always positive. This choice of $\chi_{\rm e}$ and $\varphi_{\rm
e}$ corresponds to the maximum of $P^{M(N)}$ and, from Eq.
(\ref{12}), we obtain
\begin{equation}
P^{M(N)}(\varphi_{\rm e}, \chi_{\rm e}) = P_{\rm max}^{M(N)} =
\frac{1}{2} + \sqrt{R^2+ |Q|^2} .
\label{11a}
\end{equation}
The corresponding detection state, onto which a projection has to
be performed in a measurement scheme achieving the maximum
probability, is determined by $|\mu_{\rm e}\rangle =
\cos\varphi_{\rm e} \,|\psi_1\rangle
      + {\rm e}^{i\chi_{\rm e}} \sin \varphi_{\rm e} \, |v\rangle$.

As applications of this general expression, we discuss two special
cases. First, the solution can be cast to a considerably simpler
form when the states are {\it real}. Real states have been
considered before \cite{sasaki} in a different context. In this
case the parameters of the optimum detection state, $|\mu\rangle$,
can be calculated very easily. Both $R$ and $Q$ are real, yielding
$\chi_{\rm e}=0$ if $Q\geq 0$ or $\pi$ if $Q<0$. The maximum
probability of determining correctly to which of the two
complementary subsets a state belongs is given in this case by Eq.
(\ref{11a})
with
\begin{equation}
R = \sum_{k=1}^M \eta_k \left(A_{1k}^2-\frac{1}{2}\right)
         -\sum_{k=M+1}^N \eta_k \left(A_{1k}^2-\frac{1}{2}\right) ,
\label{13}
\end{equation}
and
\begin{equation}
Q = \sum_{k=1}^M \eta_k A_{1k}\sqrt{1-A_{1k}^2}
         -\sum_{k=M+1}^N \eta_k A_{1k}\sqrt{1-A_{1k}^2},
\label{14}
\end{equation}
where, in the last step, we made use of the relation resulting from
Eq. (\ref{9}) with $\gamma_{k}=0$ and all the overlaps are assumed real.

As our second example, we consider the case of three {\it
arbitrary} but linearly dependent states, $N=3$. Choosing $M=1$
and taking $\eta_1 + \eta_2 + \eta_3 = 1$ into account in Eq.
(\ref{15}), we readily obtain
\begin{equation}
 R = \frac{1}{2} - \eta_2 |A_{12}|^2 - \eta_3 |A_{13}|^2 .
\label{18}
\end{equation}
The evaluation of $|Q|$ is greatly facilitated if we notice that
the first sum on the r.h.s. of Eq. (\ref{16}) has only one term
and
this term vanishes. A straightforward evaluation of
the remaining two terms from the second sum yields
\begin{eqnarray}
|Q|^{2} &=&  \eta_{2}^{2}|A_{12}|^{2}(1 - |A_{12}|^{2}) + \eta_{3}^{2}
|A_{13}|^{2}(1 - |A_{13}|^{2}) \nonumber \\
{}&{}& + 2 \eta_{2} \eta_{3} ({\rm Re} A_{12}
A_{23} A_{31} - |A_{12}|^{2} |A_{13}|^{2}) .
\label{19}
\end{eqnarray}
For this case the parameters of the optimum detection state
$|\mu\rangle$ can be seen to be $\chi_{\rm e} = \chi_{\rm Q}$ and
$\tan(2\varphi_{\rm e}) = |Q|/R$, with $|Q|$ and $R$
substituted from the above equations. We do not give here a more
explicit expression for $\chi_{\rm e}$ because it is slightly
involved and enters only the detection states but not the final
result for the maximum probability. Inserting the above values of
$|Q|$ and $R$ into the general expression for the optimum
probability finally gives
\begin{eqnarray}
\lefteqn {P_{\rm max}^{1(3)}  =  \frac{1}{2} +  \frac{1}{2}
 \left[ 1 - 4  \sum_{k=2}^3
       \eta_k\,(1-\eta_k)\,|\langle\psi_1|\psi_k\rangle|^2
                \right.}
                \label{24}
                \nonumber \\
 &  \left. + 8\, \eta_2 \eta_3  \,{\rm Re}\,
 (\langle\psi_1|\psi_2\rangle \,
      \langle\psi_2|\psi_3\rangle \,\langle\psi_3|\psi_1\rangle)
      \, \right]  ^{\frac{1}{2}} .
\end{eqnarray}
This expression describes the  maximum attainable probability of
correctly distinguishing the state $|\psi_1\rangle$ from the set of
states $\{|\psi_2\rangle,|\psi_3\rangle\}$.
The minimum error probability then follows as $P_{\rm
Error}^{1(3)} = 1 - P_{\rm max}^{1(3)}$. As expected, the result
is independent of the individual phase factors of the given
states, and for $\eta_3=0$ it reduces to the pioneering formula
\cite{hel} for minimum-error discrimination between only two
non-orthogonal states.

\section{Discussion}

With respect to possible applications, the question arises how the
maximum probability for getting a correct result in quantum state
filtering compares to the maximum probability for correctly
discriminating, by means of a different measurement strategy,
between all the given states individually. In the following we
shall explore this question for a variety of symmetric  states.

Let us investigate the set of three symmetric states
\begin{equation}
|\psi_k\rangle = \cos \beta\, |u_1\rangle +
    {\rm e}^{{\rm i}\frac{2\pi}{3}(k-1)}\, \sin \beta \, |u_2\rangle ,
\label{26}
\end{equation}
with $k= 1,2,3$ and $0 < \beta \leq \pi/4$, which are
assumed to occur with equal a priori probability.
Here $|u_1\rangle$ and $|u_2\rangle$ denote any
two orthonormal basis states.
Obviously the states are linearly dependent and non-orthogonal.
Due to their symmetry, the mutual overlaps are equal and we get
$4|A_{kl}|^2=4-3\sin^2(2\beta)$ if $k \neq l$,
where we again
used the abbreviation $A_{kl}=\langle \psi_k|\psi_l\rangle$.
Moreover, we obtain that $8\,{\rm Re}\,(A_{12}A_{23}A_{31})=
8-9\sin^2 (2\beta)$.
By substituting these expressions into Eq. (\ref{24}) and taking
into account that $\eta_k =1/3$,
we find the minimum error probability for quantum state
filtering with respect to the state
$|\psi_1\rangle$,
\begin{equation}
P_{\rm Error}^{1(3)} (\beta)=
\frac{1}{6} \left[ 3 -
       \sqrt{1+3\sin^2(2\beta)} \right].
\label{27}
\end{equation}
Because of the symmetry, the same expression holds for distinguishing
any other state from the remaining two states.
For comparison, we now consider individual discrimination between all
three states.
The general formula  for minimum-error discrimination
between $N$  symmetric states, derived in Ref. \cite{ban},
has been recently applied by one of us \cite{herzog}
to states of the form (\ref{26}), yielding the maximum
probability
$ P_{\rm max}^{(1,2,3)} = \frac{1}{3}(|\sin\beta|+|\cos\beta|)^2$
for correctly distinguishing each state individually.
From this result we obtain the minimum error probability
\begin{equation}
P_{\rm Error}^{(1,2,3)}(\beta)
= 1 - P_{\rm max}^{(1,2,3)}
= \frac{1}{3}\, [2 - \sin(2\beta)].
\label{28}
\end{equation}
 The ratio
$P_{\rm Error}^{1(3)}(\beta)/ P_{\rm Error}^{(1,2,3)}(\beta)$
is found to vary between 0.5 for $\beta = 0$ or $\pi/4$,
and the maximum value 0.56 for $\beta \approx \pi/12$.
When $\beta$ approaches zero, the physical
difference between the states vanishes and the respective
minimum error probabilities,
corresponding to random guessing, are twice as large
as those for $\beta=\pi/4$, when
both kinds of minimum error probabilities take their
smallest possible values. These values are
equal to 1/3 when all three states are discriminated individually,
and to 1/6 when only one of the states is distinguished.

The same values of the respective minimum error probabilities also
result for the set of equally probable real symmetric states
$|\psi_1\rangle  = |u_1\rangle$,
$|\psi_{2}\rangle = -\frac{1}{2}(|u_1\rangle
                     + \sqrt{3}|u_2\rangle)$,
and
$|\psi_{3}\rangle = -\frac{1}{2}(|u_1\rangle
                     - \sqrt{3}|u_2\rangle)$
which are known as the trine states \cite{clarke2}. For the case
that $|u_1\rangle$ and $|u_2\rangle$  refer to a single photon and
represent horizontal and vertical linear polarization,
respectively, these states have been used to verify experimentally
the theoretical result 1/3 for the minimum error probability in
individual state discrimination \cite{clarke2}. On the other hand,
from Eq. (\ref{24}) with $\eta_k = 1/3$  we easily find that the
minimum error probability for distinguishing the state
$|\psi_1\rangle$ alone is only 1/6.
By using $\tan(2\varphi_{\rm e}) = |Q|/R$ and Eq. (\ref{10}) with
$\chi=0$, the proper projection state, $|\mu_{\rm e}\rangle$, is
found to be $|u_1\rangle$. Hence the corresponding
quantum-state-filtering experiment for single photons could be
performed with the help of a polarizing beam splitter that
transmits the horizontal component and reflects the vertical one,
or vice versa, as it is immediately expected in view of the
symmetry of the problem.


In conclusion, we remark that it is straightforward to generalize
our basic equation (\ref{2}) in order to account for
discrimination between more than two subsets. However, since the
detection operators always have to resolve the identity, they
cannot be represented by projection operators onto orthogonal
states if their number is larger than the dimensionality of the
underlying Hilbert space. The measurement therefore would be a
generalized \cite{kraus} measurement in this case. The same
applies if the number of detection operators is smaller than the
number of dimensions of the Hilbert space, as it happens if, e.
g., Eq. (\ref{2}) is applied to three linearly independent states.
Finally it is interesting to relate our results to the Helstrom
bound
$P_E = \frac{1}{2} [(1- ||w_1\rho_1 - w_2\rho_2||)]$
for the minimum error probability of discriminating between two
density operators $\rho_1$ and $\rho_2$ having the a priori
probabilities $w_1$ and $w_2$, respectively. Here the symbol
$||\cdot||$ denotes the trace norm $||\sigma|| \equiv {\rm
Tr}\sqrt{\sigma^{\dagger}\sigma}$.
 After inserting
$w_1\rho_1= \sum_{k=1}^M \eta_k |\psi_k\rangle\langle\psi_k|$ and
$w_2\rho_2= \sum_{k=M+1}^N \eta_k |\psi_k\rangle\langle\psi_k|$,
the expressions ensuing from $P_E$ for the cases we are interested
in indeed confirm our results, without yielding the optimum
detection operators, however.

To summarize, we derived the measurement strategy
that minimizes the error probability for discriminating  between
two complementary subsets of a set of $N$ non-orthogonal quantum
states spanning a two-dimensional Hilbert space. The corresponding
measurement is found to be a standard von-Neumann measurement,
projecting onto two orthonormal states that have been determined
in the paper. Assuming arbitrary a priori probabilities of the $N$
linearly dependent non-orthogonal states, we obtained a general
analytical expression for the minimum error probability or,
equivalently, for the maximum probability of obtaining a correct
result. As special cases of this general result, we gave explicit
expressions for the case of $N$ real states, Eq. (\ref{11a}),
and for three arbitrary states, Eq. (\ref{24}).

\begin{acknowledgments}
J. B. wants to acknowledge the hospitality extended to him during
his stay at the Humboldt-University in Berlin. The research of J.
B. was also supported by the Office of Naval Research (Grant
Number: N00014-92J-1233) and by a grant from PSC-CUNY.
\end{acknowledgments}

\end{document}